**Resonant structures based on amorphous silicon sub-oxide doped with $Er^{3+}$ with silicon nanoclusters for an efficient emission at 1550 nm**

D. S. L. Figueira[1], D. Mustafa[2], L. R. Tessler[1] and N. C. Frateschi[1,3]

[1] Instituto de Física "Gleb Wataghin," Universidade Estadual de Campinas-UNICAMP, São Paulo, 13083-970, Brazil.

[2] Max Planck Institut für Metallforschung, Heisenbergstr 3, Stuttgart 70569, Germany and Universität Stuttgart Institut für Materialwissenschaft Heisenbergstr 3, Stuttgart 70569, Germany

[3] Center for Semiconductor Components, Universidade Estadual de Campinas-UNICAMP, São Paulo, 13083-870, Brazil

Abstract

We present a resonant approach to enhance 1550 nm emission efficiency of amorphous silicon sub-oxide doped with $Er^{3+}$ (a-$SiO_x$<Er>) layers with silicon nanoclusters (Si-NC). Two distinct techniques were combined to provide a structure that allowed increasing approximately 12 x the 1550 nm emission. First, layers of $SiO_2$ were obtained by conventional wet oxidation and a-$SiO_x$<Er> matrix was deposited by reactive RF co-sputtering. Secondly, an extra pump channel ($^4I_{15/2}$ to $^4I_{9/2}$) of $Er^{3+}$ was created due to Si-NC formation on the same a-$SiO_x$<Er> matrix via a hard annealing at 1150º C. The $SiO_2$ and the a-$SiO_x$<Er> thicknesses were designed to support resonances near the pumping wavelength (~500 nm), near the Si-NC emission (~800 nm) and near the a-$SiO_x$<Er> emission (~1550 nm) enhancing the optical pumping process.

**Indexing terms**: resonator, erbium, silicon nanocristal, sputtering.

**Introduction**

The use of silicon-based material for photonic applications is very attractive because it allow the integration of CMOS technology and photonics[1]. Numerous attempts to obtain gain or employ Si as an active medium for direct application in optoeletronic have been reported[2, 3, 4, 5, 6]. However, this task has been shown difficult to accomplish and many alternatives were recently demonstrated, such as Raman amplification[7], hybrid integration with III-V alloys[8], Si-NC formation in amorphous Si matrices[9], and doping with rare-earth materials[10, 11]. Particularly, Si doped with $Er^{3+}$ appears to be interesting to telecom applications due to the strong emission at 1550 nm[12]. One method to obtain a matrix of amorphous silicon doped with $Er^{3+}$ (a-Si<Er>) is by RF co-sputtering[13]. Particularly, in a-Si <Er>, it appears that energy transfer from the matrix to $Er^{3+}$ ions occurs[13,14] because of the excitation spectrum follows the absorption spectrum of the amorphous silicon (a-Si). Therefore, continuous pumping with a wavelength that is absorbed by the a-Si is possible. It has been proposed that the excitation mechanism of $Er^{3+}$ in a-Si is due to a defect related Auger quasi-resonant process by a Coulomb interaction between the electrons that recombine in the dangling bonds and the electron from the $Er^{3+}$ ground state – DRAE (Defect-Related Auger Excitation)[14] and by a resonant dipole-dipole interactions originated by the non-radiative recombination of the electron-hole pars in the dangling bonds[15, 16]. In both cases the electrons in $Er^{3+}$ are excited from $^4I_{15/2}$ to $^4I_{13/2}$ levels. One common way to enhance the efficiency of the emission at 1550 nm is accessing the $Er^{3+}$ transition $^4I_{15/2}$ to $^4I_{11/2}$ using an external pump at 980 nm[17]. Another way, less frequently used, would be use the transition $^4I_{15/2}$ to $^4I_{9/2}$ at 807 nm[18].

Samples of a-Si<Er> fabricated by RF co-sputtering can be oxygenated during deposition resulting in a-SiOx<Er>. These samples, when submitted to hard annealing result in the formation of Si-NC's in the a-SiOx<Er> matrices. The Si-NC's present in these matrices have emission in a broadband region between 700 nm and 900 nm, becoming excellent candidates to pump the $^4I_{15/2}$ to $^4I_{9/2}$ $Er^{3+}$ transition. However, this transition has a small absorption cross section reducing the pump efficency to level $^4I_{13/2}$ [17]. Another problem is the reduction of the 1550 nm emission efficiency due to the hard annealing required for the Si-NC formation[19]. Therefore, even though RF sputtering followed by Si-NC formation is a very simple approach to obtaining extra pumping for $Er^{3+}$, there is an intrinsic deterioration of the emission efficiency. An approach to address this issue is to create layers that provide an amplification of the emission at the wavelengths involved in the process. Also, an optimization of the hard annealing condition is paramount. If this is accomplished, these new structures are promising for the direct application in resonators as a gain media.

In this work we have first optimized the hard annealing temperature to maximize the 1550 nm emission. Subsequently, we develop a resonant structure approach to enhance the emission efficiency of a-SiOx<Er> with Si-NC's obtained by reactive RF co-sputtering. Essentially, we combine the a-SiOx<Er> layer with a $SiO_2$/Si substrate to create an optical cavity that supports the wavelengths involved in the pumping/emission process, i.e. ~500 nm, ~800 nm and ~1550 nm. Using this technique we were able to show a 12 x enhancement in 1550 nm emission power.

**Experiment and Results**

The first step in our work was to obtain a-SiO$_x$<Er> films prepared by reactive RF co-sputtering deposition on Si (100) substrates. The base pressure of the vacuum chamber was 2x10$^{-6}$ mbar and the sputtering was carried out in RF mode with the bias fixed at 1kV from circular sources in an atmosphere of 8x10$^{-3}$ mbar of argon and a constant oxygen pressure of 5.5x10$^{-5}$ mbar. We have used an ultra pure crystalline Si (99.999%) wafer mixed with metallic erbium pieces (99%) as source material for deposition. The substrate temperature was maintained at 240 °C and the deposition rate was roughly 1.5 Å/s. This technique is a direct form to obtain simultaneously a-SiO$_x$ film doped with Er$^{3+}$ unlike other reported methods requiring plasma deposition followed by Er implantation[20,21,22]. Rutherford Back Scattering (RBS) shows a 1.96x10$^{18}$ atoms/cm$^3$ Er$^{3+}$ concentration. A refractive index of n = 2.6 at 1550 nm was measured by ellipsometry. After this, four samples were submitted to two cumulative annealing procedures for the formation of the Si-NC: a) a soft annealing consisting of 1 hour at 400 °C to remove remaining Ar from the film was done in all samples[23]; b) hard annealing was done for three of the samples at 1100 °C, 1150 °C, 1200 for 60 min in a quartz tube oven in ultra pure N$_2$ atmosphere at 3.0 l/min flux for the formation of the Si-NC[19]. After the hard annealing, photoluminescence (PL) was measured to characterize the emission efficiency near 1550 nm region (for the a-SiO$_x$<Er> emission) and in the 800 nm region (for Si-NC emission). Room temperature PL was performed using the excitation of a solid state laser emitting at 532 nm and 200 mW with a spot of approximately 100 μm of radius. PL emission, perpendicular from the sample plane, was focused at the entrance slit of a 2 m spectrometer and a liquid nitrogen cooled Ge photodetector (for 1550 nm) or a Si photodetector (for 800 nm) were used at the exit slit to obtain the spectrum. Both slits were kept at 200 μm. Fig. 1a shows the emission in the 800 nm region for the four samples. Notice that the emission occurs only after the hard annealing. This is an evidence of Si-NC formation. Also the 1150 °C hard annealing temperature results in the highest emission power. Annealing at higher temperatures leads to a degradation of the emission. Fig. 2 shows a Transmission Electron Microscopy (TEM) micrography of a 4 nm Si-NC obtained after the annealing at 1150 °C . Figure 1.b. shows typical PL emission in the 1550 nm range for a sample before any heat treatment (as deposited), the sample with soft annealing and the three with soft and hard annealing. Clearly, after the soft annealing already the emission at 1550 nm is already maximized. After 1150º C, there is deterioration in the emission efficiency. Since the soft annealing process can't form the Si-NC's, one can't say if there is an enhancement in 1550 nm due to the presence of the nanoclusters. This perhaps is due to the small absorption cross-section for pumping $Er^{3+}$ at 807 nm as we mentioned above. In order to try to improve the pumping at his wavelength the resonant approach was designed as described below.

Since we use a Si substrate, the approach to obtain resonances was to use a $SiO_2$ layer obtained by direct oxidation of the substrate as the first step. Using the matrix propagation method[24] the thickness of a-$SiO_x$<Er> and of $SiO_2$ were calculated. On these simulations we consider that: a) the desired resonances must occur in the a-$SiO_x$<Er> layer; b) there are 3 wavelenghts involved in the process (external pump near 500nm, Si-NC emission near 800 nm, and a-$SiO_x$<Er> emission near 1550 nm); c) as the a-$SiO_x$<Er> layer is deposited by reactive RF co-sputtering we consider an error of 15% in the uniformity of the film thickness in this simulation. For the $SiO_2$ layers we calculated the thicknesses that support resonances near 500 nm, 800 nm, and 1550 nm. The results of the simulation are shown in figure 3. Fig. 3a, 3b and 3c show the expected maximum field intensity within the a-$SiO_x$<Er> layer for a $SiO_2$ layer of 175 nm, 250 nm and 550 nm, respectively, and in a thickness of a-$SiO_x$<Er> layer of the 600 nm. Also, we have added gray rectangles to this figure indicating the pumping/emission wavelength with associate linewidth. Three distinct situations are obtained: (a) 175 nm $SiO_2$ thickness results in resonances near 532 nm pump; (b) 250 nm $SiO_2$ thickness results in a resonance near 807nm;(c) 530 nm $SiO_2$ thickness results in three resonances, a strong resonance at 1550 nm, one near 500 nm, and a weak resonance around 800 nm.

After simulation, we repeated the a-$SiO_x$<Er> layer deposition with subsequent Si-NC formation, however using substrates with $SiO_2$ layer. The $SiO_2$ was obtained by wet oxidation of the Si wafer at a temperature of 1100 $^o$C in a flux of 1.0 l/min of $O_2$ and water vapor. We have prepared 3 samples with 600 nm of a-$SiO_x$<Er> and different $SiO_2$ thickness, all based on the simulation results. The samples were all treated with the soft annealing followed by the hard annealing at 1150 $^o$C as described above.

We repeated the photoluminescence characterization on the samples with resonant structures. Fig. 4a and 4b show the maximum PL intensity in the 800 nm and 1550 nm regions, respectively. The dashed line in both curves shows the maximum emission obtained for the previous samples without the $SiO_2$ layer. Clearly, the presence of the $SiO_2$ layer improved the measured emission intensity at both the 800 nm and 1550 nm regions. Also, we observe that the maximum emission at 1550 nm and 800 nm occurs for a $SiO_2$ thickness of 530 nm. This agrees with the simulation shown in figure 3c, since for this $SiO_2$ thickness there are three resonances: one at the pump wavelength ~500 nm, at 532nm, a small one at near 800 nm and a strong resonance at 1550nm. For the $SiO_2$ thickness of 250 nm, where a strong resonance in the 800 nm range is predicted, we see no improvement in the emission as evidenced by fig. 4a. However, this same thickness leads to a large enhancement of the 1550 nm emission as shown in fig. 4b. This was not expected, since there is no resonance at 1550 nm. Therefore, we may conclude that the resonance is increasing the 800 nm photonic density in the a-SiO$_x$<Er> layer, but these photons are being absorbed by the Er$^{3+}$ ions through the $^4$I$_{15/2}$ to $^4$I$_{9/2}$ states and being reemitted at 1550 nm. In summary, the enhancement of the 800 nm and 1550 nm for the samples with resonances near 500 nm and 1540 nm is expected since they either improve pumping indirectly through the a-Si absorption (500 nm) or they enhance the emitted photons at 1550 nm. The PL results for the samples with resonances at 800 nm can be explained by an enhanced internal photonic density at this wavelength that is optically pumping the Er$^{3+}$ ions in the matrix.

**Conclusion**

We present a technique to fabricate a-SiO$_x$<Er> films with Si-NC's using a single step of reactive RF co-sputtering and a dual thermal annealing scheme. In order to improve the emission at 1550 nm, we produced samples with built in resonant structures using SiO$_2$ films obtained by direct wet oxidation of the substrate. The structure supported resonances near the pumping wavelength (~500nm), near the Si-NC's emission (800 nm) and near Er$^{3+}$ emission wavelength (~1550 nm). Enhancement of up to 12 x was obtained in the emission at 1550 nm. Samples with resonances near 800 nm, showed also a large enhancement in 1550 nm emission possibly due to the Si-NC's optical pumping in the Er$^{3+}$ ($^4$I$_{15/2}$ to $^4$I$_{9/2}$) channel.

**Acknowledgments:** This work was supported by the Coordenação de Aperfeiçoamento de Pessoal de Ensino Superior (CAPES), the Fundação de Amparo à Pesquisa do Estado de São Paulo (FAPESP), the Centro de Pequisa em Óptica e Fotônica (CePOF), and Instituto Nacional de Ciência e Tecnologia (INCT- FOTONICON).

**Captions:**

**Fig. 1.** (a)Photoluminescence spectra in the 800 nm region for samples submitted to different temperatures; (b) Photoluminescence spectra in the 1550 nm region for samples with no annealing and with different annealing temperatures. Maximum PL intensity occurs for a sample annealed ate 400º C (open circles) and at 1150º C (solid line).

**Fig. 2.** Transmission Electron Microscopy micrograph of Si nanocluster obtained after annealing at 1150ºC for 1 hour.

**Fig. 3:** Calculated optical field intensity within the 600 nm a-SiOx<Er> layer as a function of wavelength. (a) $SiO_2$ thickness of 175 nm; (b) $SiO_2$ thickness of 250 nm; (c) $SiO_2$ thickness of 530 nm. Gray rectangles are placed approximately to represent the pumping/emission wavelength with associate linewidth.

**Fig. 4. (a)** Maximum PL intensity in the 800 nm region as a function of $SiO_2$ thickness. (b) Maximum PL intensity in the 1550 nm region as a function of $SiO_2$ thickness. The dashed line indicates the maximum PL intensity obtained for samples without the $SiO_2$ layer.

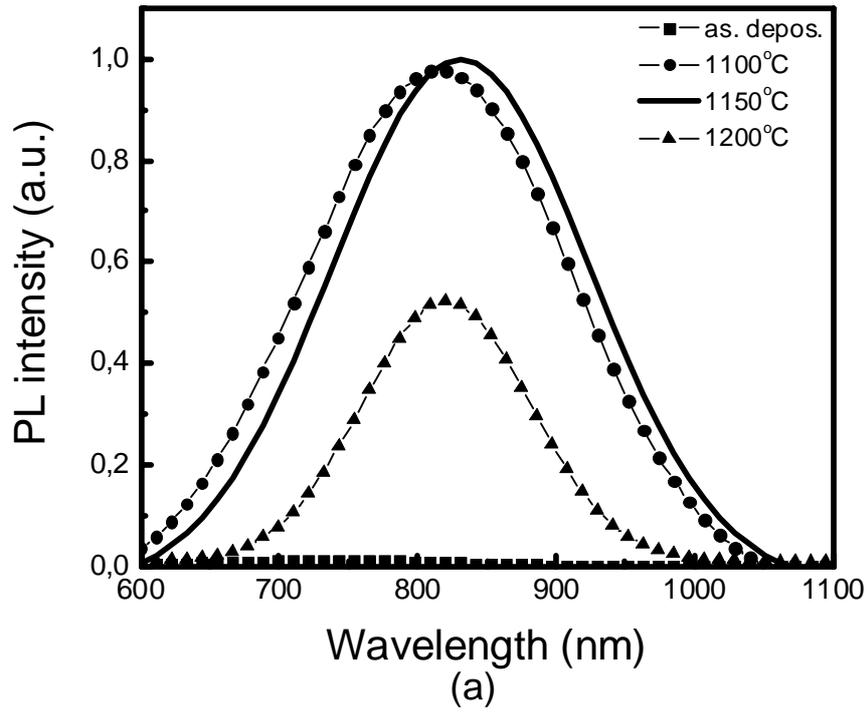

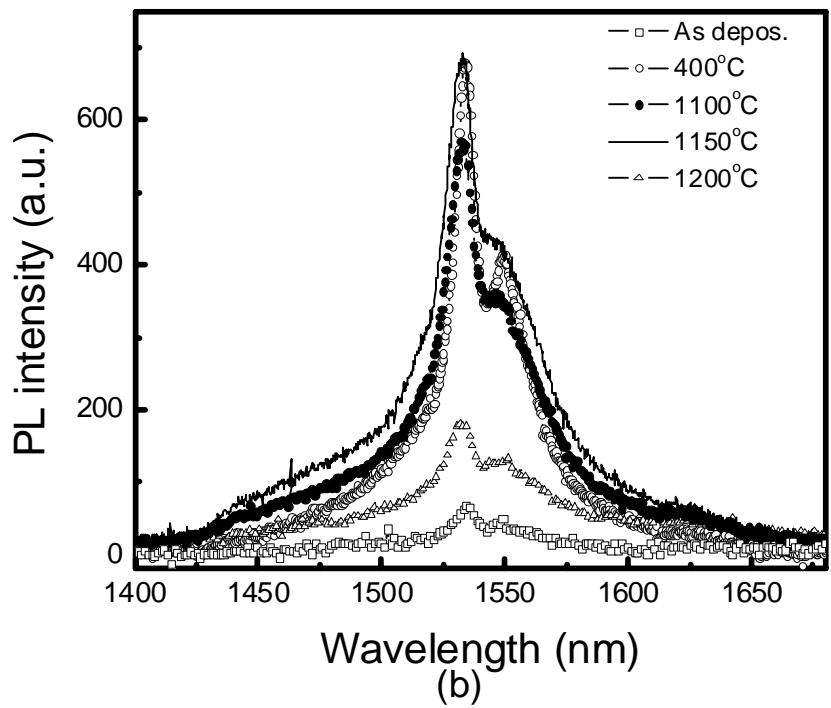

**Figure 1**

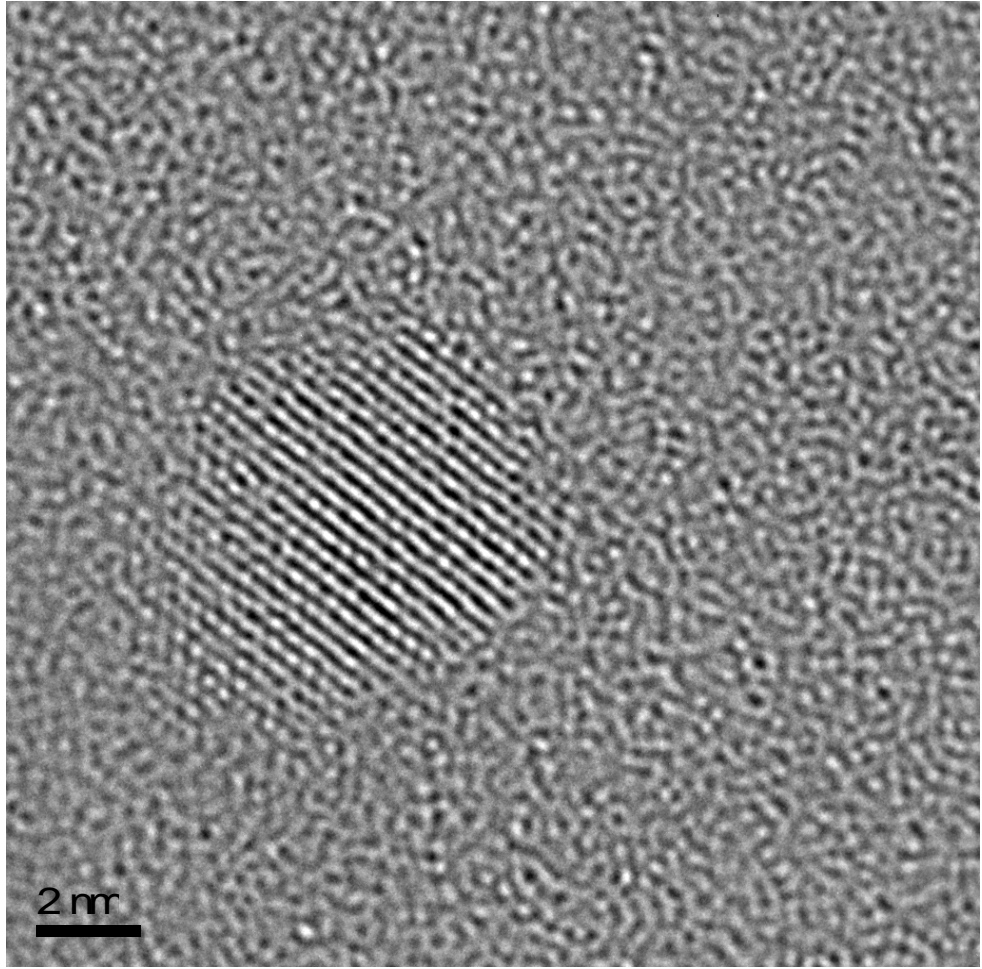

**Figure 2**

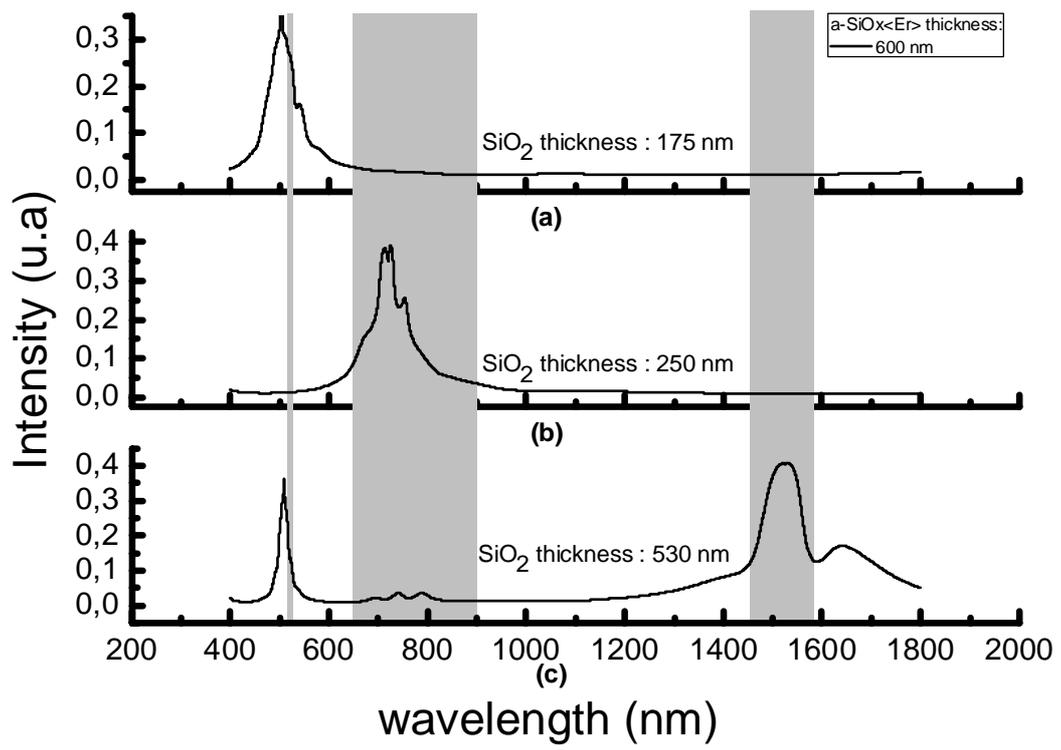

**Figure 3**

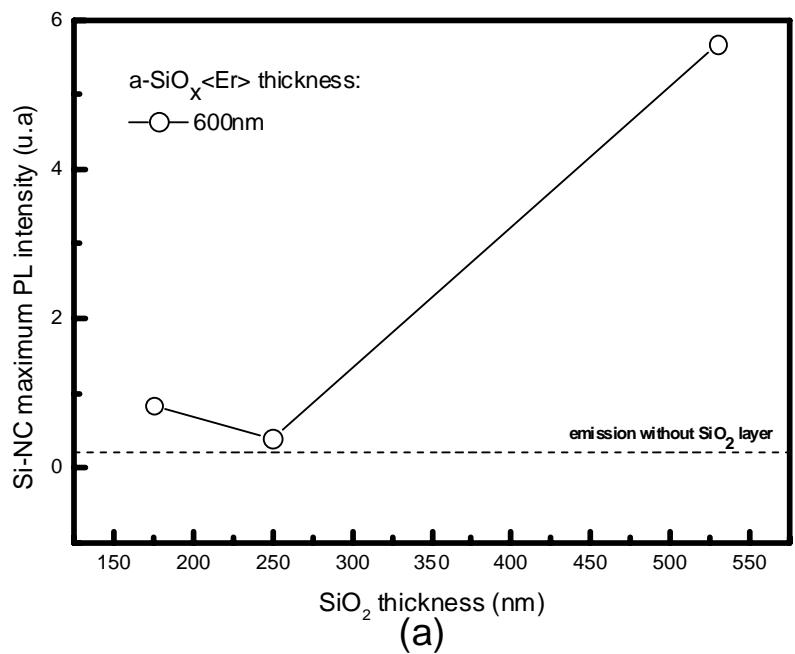

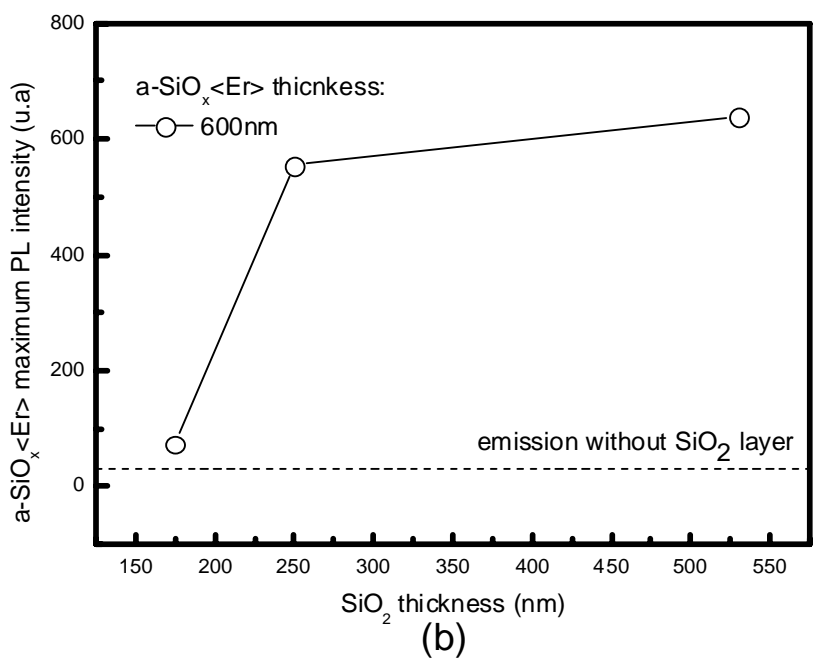

**Figure 4**